\documentstyle[aps,epsfig,prb,floats]{revtex}
\tightenlines

\def\LSCO{La$_{2-x}$Sr$_{x}$CuO$_{4}$\,}
\def\cm-1{cm$^{-1}$}

\begin{document}

\twocolumn[
\hsize\textwidth\columnwidth\hsize\csname@twocolumnfalse\endcsname

\draft

\title
{Inhomogeneous CuO$_{6}$ Tilt Distribution and Charge/Spin Correlations
in La$_{2-x-y}$Nd$_{y}$Sr$_{x}$CuO$_{4}$ around commensurate hole
concentration }

\author{
A.~Gozar$^{1,2,\dag}$, B.S.~Dennis$^{1}$, T.~Siegrist$^{1}$,
Y.~Horibe$^{1}$, and G.~Blumberg$^{1,\ddag}$
}

\address{
$^{1}$Bell Laboratories, Lucent Technologies, Murray Hill, NJ 07974 \\
$^{2}$University of Illinois at Urbana-Champaign, IL 61801-3080
}

\author{Seiki Komiya and Yoichi Ando}

\address{
Central Research Institute of Electric Power Industry, Komae,
Tokyo 201-8511, Japan}

\date{\today}
\maketitle

\newpage
\begin{abstract}

Phononic and magnetic Raman scattering are studied in
La$_{2-x-y}$Nd$_{y}$Sr$_{x}$CuO$_{4}$ with three doping
concentrations: $x \approx 1/8$, $y =
0$;  $x \approx 1/8$, $y = 0.4$; and $x = 0.01$, $y = 0$.
We observe strong disorder in the tilt pattern of the CuO$_{6}$ octahedra
in both the orthorhombic and tetragonal phases which persist down to
10~K and are coupled to bond disorder in the cation layers
around $1/8$ doping independent of Nd concentration.
The weak magnitude of existing charge/spin modulations in the Nd
doped structure does not allow us to detect the specific
Raman signatures on lattice dynamics or two-magnon scattering around
2200~\cm-1.

\end{abstract}


]
\narrowtext

The evolution of ground state properties with carrier doping and the
nature of electronic excitations in correlated systems continues to
be a
subject of intensive research \cite{SachdevMillis}.
After emerging as the mean field solution for the ground state at low
hole concentrations in doped Mott-Hubbard antiferromagnets (AF)
\cite{ZaanenPRB}, stripes have become one paradigm for understanding
the properties of high-T$_{c}$ cuprates \cite{Emery}.
Extensive studies of incommensurate neutron
scattering peaks have been performed on \LSCO (LSCO) crystals and the
development of magnetic fluctuations were related to the onset of
superconducting order \cite{Yamada}.
Transport measurements demonstrate that the charge degrees of freedom
self-organize into quasi-1D structures in these materials \cite{Ando}.
It was shown that Nd substitution in LSCO and the proximity to the
commensurate $1/8$ doping level enhances the stripe correlations
and static stripes were observed in
La$_{1.48}$Nd$_{0.4}$Sr$_{0.12}$CuO$_{4}$ (LNSCO)
\cite{Tranquada-Nature}.
X-ray diffraction data also observed weak charge order superlattice
peaks in this compound~\cite{Zimmermann}.

Lanthanum cuprates are also characterized by disorder introduced by
cation substitution and distortions of the CuO planes~\cite{Kastner}.
Above room temperature LNSCO undergoes a transition from a high
temperature tetragonal (HTT) to a low temperature orthorhombic (LTO)
phase and around 70~K a transition from the LTO to a low
temperature tetragonal (LTT) phase~\cite{Buchner,Crawford}.
The LSCO system does not enter the LTT phase but the orthorhombicity
decreases with Sr doping.
The order parameter of these transitions is the tilt angle of the
CuO$_{6}$ octahedra shown in Fig.~1.
It was noticed in LNSCO that the LTO-LTT transition takes place over
a range of temperatures and that disorder in the striped phase leads
to a glassy nature of the ground state \cite{TranquadaPRB99}.
The coexistence in L(N)SCO of several phases in a complex mixture was
suggested by transmission electron microscopy~\cite{Inoue,Horibe}.

The importance of the local CuO$_{6}$ tilt dynamics for the superconducting
properties characterized in cuprates by a short coherence length
cannot be ignored.
The stabilization of the LTT phase was observed to trace the
suppression of superconductivity in Nd doped LSCO~\cite{Crawford} and
also in the related La$_{2-x}$Ba$_{x}$CuO$_{4}$ compound~\cite{Axe}.
A critical value of the tilt angle was associated with the stabilization
of magnetic against superconducting order~\cite{Buchner}.
Rapid suppression of superconductivity, similar to that due to Cu
substitution by non-magnetic impurities, was observed with increasing
the cation radius variance~\cite{McAllister}.
By providing information about the energy, scattering width,
intensity as well as symmetry of the excitations, Raman scattering
has been powerful in studying the structural, electronic, and magnetic
properties of cuprates
\cite{SugaiPRB89,Weber,Girsh96,SugaiPRB90}.
Optical phonons can be used as local probes of fast changes in the
charge distribution and magnetic Raman scattering provides
information about local AF correlations \cite{Girsh96,Sriram}.
This was the case in related La$_{2-x}$Sr$_{x}$NiO$_{4 + \delta}$
compounds\cite{Girsh98,Pashkevich} where the existence of fully
developed striped phases is well established \cite{Chen}.

Our study provides direct spectroscopic information about the LTO-LTT
transition in LNSCO and local deviations from the average structure
existent in Nd doped and Nd free LSCO structures.
Strong local disorder in the octahedra tilts persists down to T~=~10~K and
reveals strong disorder in the cation-oxygen layers.
The distinct Raman signatures accompanying a transition to a state
with deep spin/charge modulations are not observed in the temperature
dependence of the two-magnon (2M) scattering around 2200~\cm-1 and
the $c$-axis polarized phonons below 500~\cm-1.

Single crystals of LNSCO, $x = 0.01$ and $x \approx 1/8$ LSCO were grown
by the traveling-solvent floating-zone technique.
Raman spectra were taken from the $(ac)$ and $(ab)$ faces of $x \approx 1/8$
L(N)SCO and from the $(a'c)$ face of $x = 0.01$ LSCO crystals as determined
by X-ray diffraction.
Unless otherwise stated the laser excitation energy was $\omega_{in}
= 1.92$~eV.
\begin{figure}[t]
\centerline{
\epsfig{figure=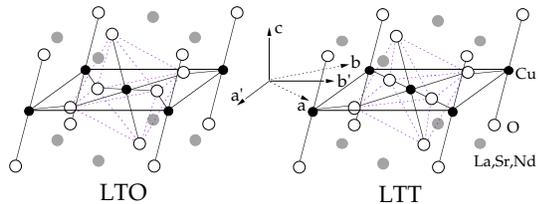,width=70mm}
}
\caption{
Cartoon showing the buckling of the Cu-O plane in the LTO and LTT
phases and adjacent cation-oxygen layers.
The $a'$, $b'$ axes are rotated by~45$^{\circ}$ with respect to the
$a$ and $b$ coordinates.
In a rigid approximation the CuO$_{6}$ octahedra (shown by dotted
lines) are tilted around the $a'$ axis in the LTO and $a$ axis
in the LTT phase.
}
\label{Fig.1}
\end{figure}

Raman spectra taken in the $(ca)$ geometry provide direct information
about tetragonal to orthorhombic distortions.
In this polarization we probe phononic modes with B$_{2g}$ and
B$_{3g}$ symmetries in the LTO phase which become degenerate with
E$_{g}$ symmetry in the LTT phase.
The temperature dependence of the modes around 250~\cm-1
corresponding to the apical oxygen vibrations parallel to the Cu-O plane
in LNSCO is shown in Fig.~2.
We observe a broad peak around 245~\cm-1 at room temperature which,
with cooling, becomes resolved into two components.
A new central peak can be seen at 50~K around 248~\cm-1 which gains
spectral weight as the temperature is decreased to 10~K.
While the total integrated intensity of the modes remains
constant we observe a redistribution of
spectral weight among the three modes as a function of temperature.
The split components become weaker but can still be seen as
'orthorhombic satellites' of the central peak down to 10~K.
The coalescence of the features into the 248~\cm-1 mode signals the
recurrence of a phase with tetragonal symmetry which should be the
expected LTT phase of LNSCO.
However, the finite residual intensity of the satellites appearing on
the tail of the broad central peak shows an incompletely developed 
LTT phase and that even at 10~K there exists about 7\% LTO phase 
(determined from the
relative ratio of phonon intensities).

\begin{figure}[t]
\centerline{
\epsfig{figure=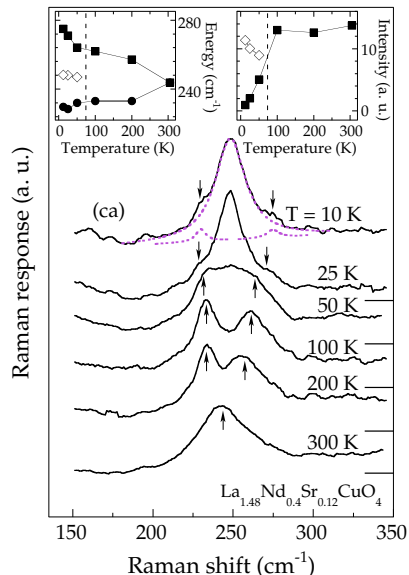,width=55mm}
}
\caption{
Raman spectra of LNSCO with light polarization of the incoming and
outgoing photons along the $c$ and $a$ axes, show the temperature
evolution of the apical oxygen vibrations.
The arrows trace the peak positions of the orthorhombic components
(a 3 Lorentzian fit is shown for 10~K data).
Insets: Temperature dependence of the energy and integrated intensity
of the oxygen modes in the LTO (filled symbols) and LTT (empty
diamonds) phases.
The vertical dashed lines in the insets mark the LTO-LTT transition.
}
\label{Fig.2}
\end{figure}
Raman data in $(cc)$ polarization is well suited for the study of
lattice dynamics due to weaker coupling to underlying electronic
excitations.
Group theory predicts five fully symmetric modes at ${\bf k} = 0$ in
each of the LTO and LTT phases and this is what we observe in Fig.~3
for L(N)SCO crystals in $(cc)$ geometry.
These modes correspond to: the tilt of the CuO$_{6}$ octahedra (mode
A), the vibration of La/Sr/Nd atoms in the direction imposed by the
CuO$_{6}$ tilt (mode B) and along the $c$ axis (mode C), the $c$ axis
vibrations of the in-plane O (mode D) and apical O atoms (mode E)
\cite{SugaiPRB89,Weber} (the modes C and E are the two fully 
symmetric modes predicted by group theory for the HTT phase).
The above qualitative description indicates that we could expect a
strong coupling between the lowest energy modes (A and B).
These two modes can be distinguished because they remain much
broader even at the lowest temperature in comparison with the the
modes C, D and E which harden and sharpen smoothly through the
LTO-LTT transition.
The temperature variation of modes C and E inherited from the HTT 
phase is not as pronounced.
The large variation in the energy and width of mode A above the
transition, the softening below 70~K, as well as its energy
around 110~\cm-1 in agreement with neutron scattering
studies~\cite{Thurston} show that this mode corresponding to the
octahedra tilt is the soft mode of the structural changes.
The apparently smooth decrease in energy in the LTT phase is peculiar
because this space group is not a subgroup of the LTO group and as a
result a true LTO-LTT transition is expected to be of first order.
Although unresolved due to broadening effects, the large width of
mode A around 70~K shows the coexistence of the LTO and LTT tilts,
the latter appearing as a result of folding of the LTO $Z$-point to
the $\Gamma$-point of the LTT phase as observed for instance in
La$_{2}$NiO$_{4}$~\cite{Burns}.

\begin{figure}[t]
\centerline{
\epsfig{figure=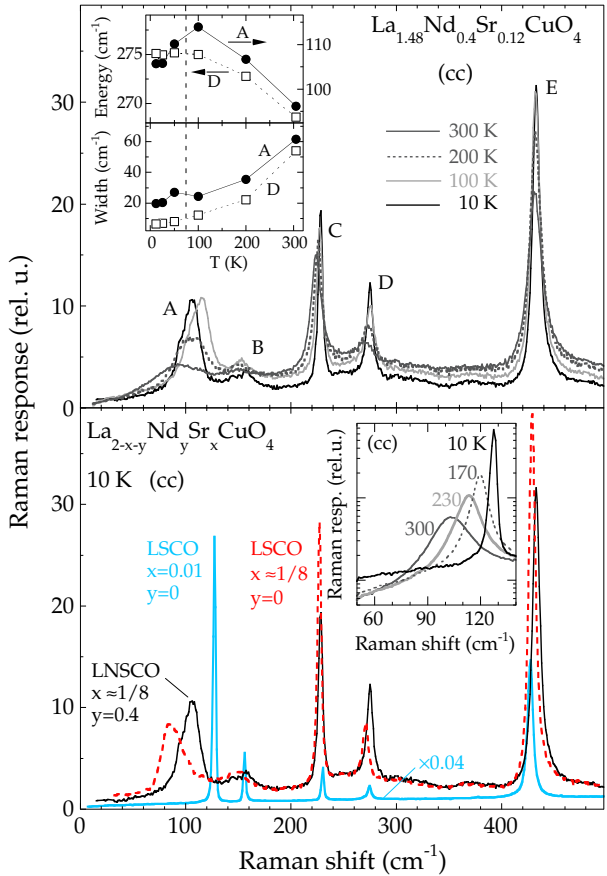,width=80mm}
}
\caption{
Top: Temperature dependence in $(cc)$ polarization of the 5 fully
symmetric modes in LNSCO.
Variation with temperature of the energy (upper inset) and the full
width at half maximum (lower inset) of mode A (solid circles) and D
(empty squares).
Vertical lines in the insets correspond to the structural LTO-LTT
transition.
Bottom: 10~K Raman spectra in $(cc)$ configuration for L(N)SCO.
Inset: Temperature dependence of the intensity of mode A in $x = 0.01$
LSCO.
Note the semi-log scale.
}
\label{Fig.3}
\end{figure}

To understand the surprising behavior of the tilt pattern as
reflected in the LNSCO phononic data a comparison with different
materials from the same class is necessary.
In the lower panel of Fig.~3 we show the 10~K $(cc)$ polarized Raman
spectrum of L(N)SCO.
In particular, mode A at 128~\cm-1 for $x = 0.01$ LSCO has a
full width at half maximum (FWHM) of 2.5~\cm-1.
Note in the inset the strongly temperature dependent
intensity and width which is a characteristic of a soft mode.
For $x \approx 1/8$~LSCO the same phonon is around 85~\cm-1 and
its FWHM of about 23.5~\cm-1 is larger than the width of mode A
in the Nd doped crystal where it is slightly below 20~\cm-1.
Comparison of the relative phononic widths for both the soft mode and the
$c$-axis vibrations shows that Nd doping of LSCO crystals and the
closer proximity to the T$^{\prime}$ phase induced by Nd doping in the
La$_{2}$CuO$_{4}$ structure cannot be responsible for the large
observed broadening effects.
Intrinsic phonon anharmonicity would lead to a broad mode A in
$x = 0.01$~LSCO which is not the case.
The tilt disorder across twin domains cannot be the cause of such
dramatic effects because the volume fraction occupied by these
boundaries is expected
to be very small \cite{Inoue,Horibe}.
The 7\% relative ratio of the orthorhombic satellites to the central
peak in Fig.~2 may be consistent with such a small contribution.
If the satellites are indeed due to twinning effects the data show
that at 10~K the larger LTT domains are separated by regions of pure
LTO tilt.
The absence of the broadening effects on the vibrations along the
$c$-axis points towards an 'anisotropic' disorder relating primarily
to bond randomness along directions parallel to the CuO planes.
\begin{figure}[t]
\centerline{
\epsfig{figure=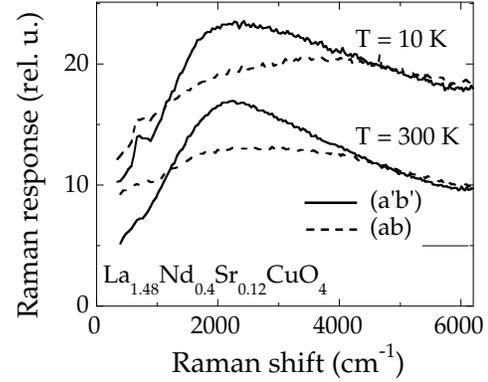,width=65mm}
}
\caption{
High energy two-magnon Raman scattering spectra in $(ab)$ and
$(a'b')$ polarizations at 300 and 10~K taken with
$\omega_{in}$~=~3.05~eV excitation energy.
The 10~K spectra are offset.
}
\label{Fig.4}
\end{figure}

Could the spin-lattice coupling or the interaction with the
stripe-ordered carriers in CuO planes be the main cause
of broadening?
Stripe correlations are enhanced in LNSCO which displays however a
smaller width of mode A.
Also, it is not clear how the $c$-axis polarizability would couple to
quasi-2D magnetic properties of the CuO planes and why only modes
A and B are affected by this interaction.
In this sense one expects the in-plane atomic movements to be more 
sensitive to stripe ordering but we see no similar effects on mode D.
Although less probable, spin-lattice induced broadening cannot be
completely ruled out and the answer to this question lies in a Sr
doping dependence of the $(cc)$ polarized spectra.
Our data can be reconciled however with recent studies of local
structure in Nd free and Nd doped L(N)SCO system~\cite{Haskel,Han}.
Model analysis of the pair distribution function from X-ray absorption
fine structure suggest that in this material class the average
structure determined by diffraction is different than the local pattern
which is characterized by disorder in the CuO$_{6}$ tilt direction and
magnitude~\cite{Haskel,Han}.
The Raman data shown in Fig.~3 is a direct spectroscopic evidence
that the L(N)SCO system is characterized by strong disorder in the
cation layers.
The locally fluctuating octahedra tilt is responsible for the
observed effects.

Information about the relative magnitude of charge disproportionation
in LNSCO can be gained by comparison with Raman spectra in compounds
where charge modulations are well established.
New Raman active modes have been observed in $x = 0.33$ and 0.225
La$_{2-x}$Sr$_{x}$NiO$_{4}$ by Raman scattering
\cite{Girsh98,Pashkevich}.
Lowering of the crystal symmetry at the stripe ordering transition
gives rise to folding of the Brillouin zone (BZ) and the appearance
of extra ${\bf k} = 0$ phononic modes.
Charge localization creates non-equivalent Ni sites generating phonon
'splitting'.
The $c$ axis stretching modes corresponding to La and apical oxygens
are split by 14 and 30~\cm-1 respectively \cite{Girsh98}.
Within the 3~\cm-1 resolution imposed by the phononic widths we do
not observe such a split in our spectra.
The ratio of the integrated intensities of the split oxygen modes in
Ref.~\cite{Girsh98} is about the same as the ratio of doped versus
undoped Ni sites.
If we assume the same relation holds for the case of cuprates, a
factor of $\approx 1/8$ in split phononic intensity should have been seen
in the spectra~\cite{footnote1}.
We conclude that any charge ordering taking place in our
case is much weaker than in the related compounds referred to above.
This does not contradict the X-ray diffraction data \cite{Zimmermann}
which estimated a factor of 10$^{2}$ relative difference between the
magnitude
of charge modulations in cuprates and nickelates.

For 2D magnetic square lattices a 2M peak in the $(a'b')$
geometry is expected \cite{Girsh96,Sriram}.
Fig.~4 shows 2M scattering around 2200~\cm-1 at 300 and 10~K taken
with the resonant $\omega_{in}$~=~3.05~eV incident frequency.
As in other tetragonal 2D AFs we observe the spin pair excitations in
the expected geometry which corresponds to B$_{1g}$ symmetry in the
HTT phase.
Renormalization of the 2M excitation energy occurs across the spin
ordering transition in the striped phase of
La$_{2-x}$Sr$_{x}$NiO$_{4}$.
The 2M Raman band around 1650~\cm-1 characteristic of the undoped
case \cite{SugaiPRB90} is not seen while the spectra exhibit lower
energy peaks  below the magnetic ordering temperature as a result of
the new spin exchange channels within and across the antiphase AF
domains \cite{Girsh98,Pashkevich}.
Our data in LNSCO shows only slight changes from 300 to 10~K
emphasizing
weak local spin modulations.

The differences we observe between cuprates and
nickelates can be related to the much stronger carrier
self-confinement in the latter \cite{Anisimov}.
It has also been shown \cite{McQueeney} that anomalies in phonon
dispersions occur in \LSCO at points in the BZ commensurate with
charge ordering wavevectors inferred from neutron scattering studies.
But as discussed, the charge modulation in Nd doped structures,
where the stripe correlations were shown to be stabilized, is too
weak to produce observable changes (at ${\bf k} = 0$) in the
lattice unit cell on time scales longer than the phononic frequencies.
The number of phononic modes we observe can be explained solely in
terms of LTO/LTT distortions.
Our data, however, does not contradict the possible existence of
charge modulations in the Cu-O plane. In fact, the dynamics in the
cation-O layers and the magnitude of octahedra tilt disorder affects 
the carrier distribution and our Raman results impose constraints on 
the magnitude of the charge modulations.

We studied lattice dynamics and magnetic scattering in stripe ordered
L(N)SCO crystals.
The transition to the LTT phase in LNSCO is not complete as shown
by the behavior of the split oxygen modes around 250~\cm-1.
The large widths of the soft modes demonstrate a locally fluctuating
pattern of the CuO$_{6}$ octahedra tilts.
This disorder effect is even stronger in the Nd free LSCO compound at
$x \approx 1/8$ while it is not present for $x = 0.01$ Sr
concentration.
Quantitative comparison with lanthanum-nickelate compounds shows that
the magnitude of the existent spin/charge disproportionations in
L(N)SCO are relatively weak and that, as in other cuprates, the two-magnon
excitation can be described within the usual high temperature
tetragonal structure.


\vspace{-5mm}
\begin{references}

\vspace{-15mm}

\bibitem[\dag]{byline} E-mail: gozar@lucent.com
\bibitem[\ddag]{byline} E-mail: girsh@bell-labs.com

\bibitem{SachdevMillis}
S. Sachdev, Science {\bf 288}, 475 (2000); J. Orenstein, and A.J.
Millis, \emph{ibid.} {\bf 288}, 468 (2000) and references therein.

\bibitem{ZaanenPRB}
J. Zaanen, and O. Gunnarsson, Phys. Rev. B {\bf 40}, 7391 (1989).

\bibitem{Emery}
V.J. Emery, and S.A. Kivelson, Nature, {\bf 374}, 434 (1995);
V.J. Emery, S.A. Kivelson, and O. Zachar, Phys. Rev. B {\bf 56}, 6120
(1997).

\bibitem{Yamada}
K. Yamada {\em et al.}, Phys. Rev. B {\bf 57}, 6165 (1998).

\bibitem{Ando}
Y. Ando {\em et al.}, Phys. Rev. Lett. {\bf 87}, 017001 (2001); Y.
Ando {\em et al.}, \emph{ibid.} {\bf 88}, 137005 (2002).

\bibitem{Tranquada-Nature}
J.M. Tranquada {\em et al.}, Nature {\bf 375}, 561 (1995); J. M.
Tranquada {\em et al.}, Phys. Rev. B {\bf 54}, 7489 (1996).

\bibitem{Zimmermann}
M.V. Zimmermann {\em et al.}, Europhys. Lett. {\bf 41}, 629 (1998).

\bibitem{Kastner}
M.A. Kastner {\em et al.}, Rev. Mod. Phys. {\bf 70}, 897 (1998) and
references therein.

\bibitem{Buchner}
B. B\"{u}chner {\em et al.}, Phys. Rev. Lett. {\bf 73}, 1841 (1994).

\bibitem{Crawford}
M.K. Crawford {\em et al.}, Phys. Rev. B. {\bf 44}, R7749 (1991).

\bibitem{TranquadaPRB99}
A.R. Moodenbaugh {\em et al.}, Phys. Rev. B. {\bf 58}, 9549 (1998);
J.M. Tranquada, N. Ichikawa, and S. Uchida, Phys. Rev. B {\bf 59},
14712 (1999) .

\bibitem{Inoue}
Y. Inoue, Y. Horibe, and Y. Koyama, Phys. Rev. B {\bf 56}, 14176
(1997).

\bibitem{Horibe}
Y. Horibe, Y. Inoue, and Y. Koyama, Phys. Rev. B {\bf 61}, 11922
(2000).

\bibitem{Axe}
J.D. Axe {\em et al.}, Phys. Rev. Lett. {\bf 62}, 2751 (1989).

\bibitem{McAllister}
Judith A. McAllister, and J. Paul Attfield, Phys. Rev. Lett. {\bf
83}, 3289 (1999).

\bibitem{SugaiPRB89}
S. Sugai {\em et al.}, Phys. Rev. B {\bf 39}, 4306 (1989).

\bibitem{Weber}
W.H. Weber {\em et al.}, Phys. Rev. B {\bf 38}, 917 (1988).

\bibitem{Girsh96}
G. Blumberg {\em et al.}, Phys. Rev. B {\bf 53}, R11930 (1996).

\bibitem{SugaiPRB90}
S. Sugai {\em et al.}, Phys. Rev. B {\bf 42}, R1045 (1990).

\bibitem{Sriram}
B.S. Shastry, and B.I. Shraiman, Phys. Rev. Lett. {\bf 65}, 1068
(1990).

\bibitem{Girsh98}
G. Blumberg, M.V. Klein, and S-W. Cheong, Phys. Rev. Lett. {\bf 80},
564 (1998); K. Yamamoto {\em et al.}, Phys. Rev. Lett. {\bf 80}, 1493
(1998).

\bibitem{Pashkevich}
Yu.G. Pashkevich {\em et al.}, Phys. Rev. Lett. {\bf 84}, 3919
(2000).

\bibitem{Chen}
C.H. Chen, S-W. Cheong, and A.S. Cooper, Phys. Rev. Lett. {\bf
71}, 2461 (1993).

\bibitem{Thurston}
T.R. Thurston {\em et al.}, Phys. Rev. B {\bf 39}, 4327 (1989).

\bibitem{Burns}
G. Burns {\em et al.}, Phys. Rev. B {\bf 42}, R10777 (1990).

\bibitem{Haskel}
D. Haskel {\em et al.}, Phys. Rev. Lett. {\bf 76}, 439 (1996).

\bibitem{Han}
S.-W. Han {\em et al.}, Phys. Rev. B {\bf 66}, 094101 (2002).

\bibitem{footnote1}
This argument has to take into account that different electron-phonon
coupling might change this proportionality relation.

\bibitem{Anisimov}
V.I. Anisimov {\em et al.}, Phys. Rev. Lett. {\bf 68}, 345 (1992).

\bibitem{McQueeney}
R.J. McQueeney {\em et al.}, Phys. Rev. Lett. {\bf 87}, 077001
(2001).




\end{references}
\end{document}